\documentclass[twocolumn]{bmcart}

\usepackage[utf8]{inputenc} 

\usepackage[procnumbered,linesnumbered,ruled,vlined]{algorithm2e}
\usepackage{booktabs}
\usepackage{amsmath}
\usepackage{graphicx}

\newcommand{\tabitem}{~~\llap{\textbullet}~~}
\newcommand{\revA}[1]{\textcolor{black}{#1}}
\graphicspath{ {./imgs/} }
\startlocaldefs
\endlocaldefs

\begin{document}

\begin{frontmatter}

\begin{fmbox}
\dochead{Research}


\title{Efficient logging and querying for Blockchain-based cross-site genomic dataset access audit}


\author[
   addressref={aff1},                   
   corref={aff1},                       
   email={sma30@emory.edu}   
]{Shuaicheng Ma}
\author[
addressref={aff2},      
email={yang@i.kyoto-u.ac.jp}   
]{Yang Cao}
\author[
   addressref={aff1},
   email={lxiong@emory.edu}
]{Li Xiong}


\address[id=aff1]{
  \orgname{Department of Computer Science, Emory University}, 
  \street{400 Dowman Dr},                     %
  \city{Atlanta},                              
  \state{GA},
  \cny{USA}                                    
}
\address[id=aff2]{%
  \orgname{Department of Social Informatics, Kyoto University},
  \city{Kyoto},
  \cny{Japan}
}


\begin{artnotes}
\end{artnotes}



\begin{abstractbox}

\begin{abstract} 
\parttitle{Background} 
Genomic data have been collected by different institutions and companies and need to be shared for broader use.
In a cross-site genomic data sharing system, a secure and transparent access control audit module plays an essential role in ensuring the accountability.
A centralized access log audit system is vulnerable to the single point of attack and also lack transparency since the log could be tampered by a malicious system administrator or internal adversaries.
Several studies have proposed blockchain-based access audit to solve this problem but without considering the efficiency of the audit queries.
The 2018 iDASH competition first track provides us with an opportunity to design efficient logging and querying system for cross-site genomic dataset access audit.  
We designed a blockchain-based log system which can provide a light-weight and widely compatible module for existing blockchain platforms.
The submitted solution won the third place of the competition.
In this paper, we report the technical details in our system.

\parttitle{Methods} 
We present two methods: baseline method and enhanced method. We started with the baseline method and then adjusted our implementation based on the competition evaluation criteria and characteristics of the log system. To overcome obstacles of indexing on the immutable Blockchain system, we designed a hierarchical timestamp structure which supports efficient range queries on the timestamp field. 

\parttitle{Results} 
We implemented our methods in Python3, tested the scalability, and compared the performance using the test data supplied by competition organizer. 
We successfully boosted the log retrieval speed for complex AND queries that contain multiple predicates. 
For the range query, we boosted the speed for at least one order of magnitude. 
The storage usage is reduced by 25\%.

\parttitle{Conclusion} 
We demonstrate that Blockchain can be used to build a time and space efficient log and query genomic dataset audit trail. 
Therefore, it provides a promising solution for sharing genomic data with accountability requirement across multiple sites.    

\end{abstract}




\end{abstractbox}
\end{fmbox}

\end{frontmatter}



\section*{Background}

With the rapid development of biomedical and computational technologies, a large amount of genomic data sets have been collected and analyzed in national and international projects  such as Human Genome Project \cite{collins_human_2003} , the HapMap project \cite{consortium_international_2003} and the Genotype-Tissue Expression (GTEx) project \cite{lonsdale_genotype-tissue_2013}, which yielded invaluable research data and extended the boundary of human knowledge. 
Thanks to the advance of computer technology,  the cost of genomic testing is dropping exponentially.
Nowadays, the testing price ranges from under \$100 to more than \$2,000, depending on the nature and complexity of the test \cite{wetterstrand_dna_2013}.
One can test her gene easily and cheaply by using services from  DNA-testing companies such as Ancestry and 23andMe.
Given the above, genomic data sets have been scattered around the world in different institutions and companies. 
On the other hand, the potential business value of genomic data and privacy concerns \cite{malin_biomedical_2013, gkoulalas-divanis_publishing_2014, naveed_privacy_2015} hinder the sharing of cross-sites genomic data.
Notably, the General Data Protection Regulation (GDPR) restricts the exchange of personal data.
Under GDPR, such sensitive data only could be accessed after obtaining the consent of data subjects (i.e., the one who owns the data) and providing accountability audit. 
This requires that any cross-site genomic data sharing system should be equipped with a secure and transparent access control module.

Blockchain technology has received increasing attention because it provides a new paradigm of value exchange.
Although it stems from cryptocurrency, many studies have investigated the adoption of blockchain in different application scenarios beyond financial domain that typically involve multiple parties with  conflict of interests such as personal data sharing \cite{zyskind_decentralizing_2015, zyskind_enigma:_2015, froelicher_unlynx:_2017}, supply chain \cite{hackius_blockchain_2017, garcia-banuelos_optimized_2017, abeyratne_blockchain_2016},  identity management \cite{azouvi_who_2017, yasin_online_2016} and medical data management \cite{kuo_modelchain:_2018, yue_healthcare_2016, xia_medshare:_2017, azaria_medrec:_2016, genestier2017blockchain, choudhury2018enforcing, li_blockchain-based_2019}. 
They show that using blockchain technology can reduce friction and increase transparency. 
A blockchain system has several notable features: decentralization, immutability and transparency.
These are achieved by cryptographic hash, consensus algorithm and many other innovations  from  previously unrelated fields such as cryptography and distributed computation \cite{narayanan_bitcoins_2017}.
Due to the space limitation, we do not introduce more details of blockchain technologies and refer interested readers to surveys on blockchain \cite{kuo_blockchain_2017, underwood_blockchain_2016, sun_blockchain-based_2016, worner_bitcoin_2016, bonneau_sok:_2015, tschorsch_bitcoin_2016, pilkington_blockchain_2016, zheng_overview_2017}.

Several studies investigated blockchain-based access log audit \cite{suzuki_blockchain_2017, castaldo_blockchain-based_2018, liang_provchain:_2017} (we introduce them in the next section). 
They focus on how to achieve the immutability of the log.
However, none of them investigated the efficiency of logging and querying for a blockchain system at the application layer. 
On the other hand, a few recent studies \cite{dinh_untangling_2017, dinh_blockbench_2017, wang_forkbase_2018, xu_cub_2018} from database community consider a blockchain  system as a distributed database, and attempt to improve the performance of such system by exploring new designs of  bottom layers  (such as storage or transaction processing) of the system.
However, without considering the application characteristics, such modifications on the back-end engine of the system may not have the desired performance improvement on every application or even cause unexpected side effects. 

The 2018 iDASH competition first track, ``\textit{Blockchain-based immutable logging and querying for cross-site genomic dataset access audit trail}", provides us with an opportunity to explore a light-weight and widely compatible  access audit module for existing blockchain platforms.
Our submitted solution won the third place of the competition.
In this paper, we report the system design and technical details in our solution.

\subsection*{The competition task \cite{idash}}
The goal of iDASH competition 2018 first track is to develop blockchain-based ledgering solutions to log and query the user activities of accessing genomic datasets across multiple sites.
Concretely, given a genomic data access log file in which each entry includes seven attributes including $\mathsf{Timestamp}$, $\mathsf{Node}$, $\mathsf{ID}$, $\mathsf{Ref-ID}$,  $\mathsf{User}$,  $\mathsf{Activity}$,  $\mathsf{Resource}$, the task is to design a time/space efficient data structure and mechanisms to store and retrieve the logs based on Multichain version 1.0.4 \cite{multichain-whitepaper}.

\paragraph{Competition setup and requirement.}
It is required that each entry in the data access log must be saved individually as one transaction (i.e., participants cannot save the entire file in just one transaction), and all log data and intermediate data (such as index or cache) must be saved on-chain (no off-chain data storage allowed). 
Competition participants can determine how to represent and store each log entry in transactions. It does not need to be a plain text copy of the log entry. Also, the query implementation should allow a user to search the log using any field of one log entry (i.e., node, id, user, resource, activity, timestamp, and a “reference id” referring to the id of the original resource request), any ``AND” combination (e.g., node AND id AND user AND resource), and any timestamp range (e.g., from 1522000002418 to 1522000011441) using a command-line interface. Also, the user should be able to sort the returning results in ascending/descending order with any field (e.g., timestamp). There will be 4 nodes in the blockchain network, and 4 log files to be stored. Users should be able to query the data from any of the 4 sites. Participants can implement any algorithms to store, retrieve, and present the log data correctly and efficiently.

\paragraph{Evaluation Criteria.}
The logging/querying system needs to demonstrate good performance (i.e., accurate query results) by using a testing dataset, which is different from the one provided for the participants. 
The speed, storage/memory cost, and scalability of each solution will be evaluated. 
The competition organizer used the binary version of Multichain 1.0.4 on 64-bit Ubuntu 14.04 with the default parameters as the test bed for fairness. 
No modification of the underlying Multichain source code is allowed. 
The submitted executable binaries should be non-interactive (i.e., depend only on parameters with no input required while it works), and should contain a readme file to specify the parameters. 
The organizer tested all submissions using 4 virtual machines, each with 2-Core CPU, 8GB RAMs and 100GB storage.

\begin{figure*}[h!]
	\centering
	\caption{Overview of the logging system.}
	\includegraphics[width=15cm]{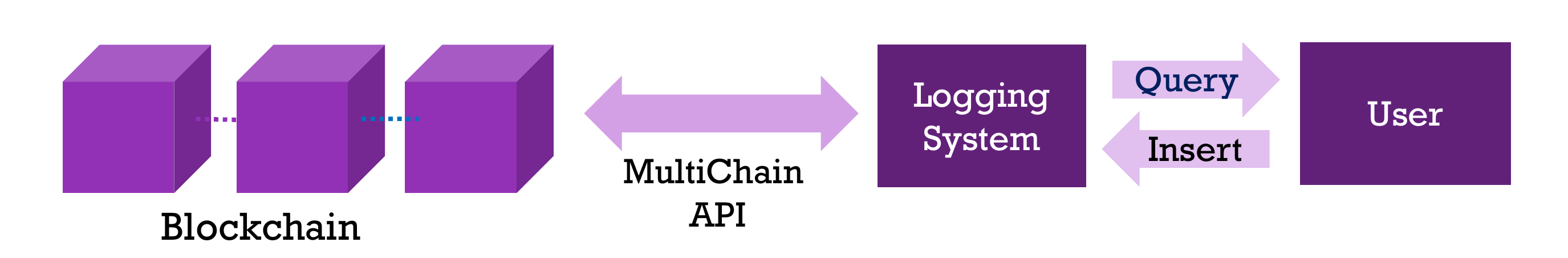}
	\label{fig:overview}
\end{figure*}

\subsection*{Related work}

The closest line of work to this competition is blockchain-based access log audit.
Suzuki et al., \cite{suzuki_blockchain_2017} proposed a method using blockchain as an audit-able communication channel.
This study is motivated by a similar problem studied in this paper: in a client-server system, the logging on either server-side or client-side does not provide strict means of auditing, because the host of the logging system could tamper the log.
They implemented a proof-of-concept system on top of Bitcoin by encoding the messages (i.e., API calls from clients and Replies from the server) between clients and the server into the transactions of bitcoin.
Since the transactions are publicly available, they can be retrieved and verified by an auditor as needed.
The proposed system is easy to use and convenient for a client-server system.  
However, answering the audit query using the proposed system may be time-consuming, especially for a large-scale system serving millions of clients, as each reply is returned in the form of a bitcoin transaction. The maximum transaction processing capacity of bitcoin  is estimated between 3.3 and 7 transactions per second\cite{croman_scaling_2016}.

Castaldo et al., \cite{castaldo_blockchain-based_2018} implemented a blockchain-based tamper-proof audit mechanism for OpenNCP (Open National Contact Points) \cite{fonseca_openncp:_2015}, which is a system for exchanging  eHealth data between countries in Europe.
The idea is similar to the one proposed in \cite{suzuki_blockchain_2017}, but dealing with data exchange instead of answering queries. 
They also encode the data that need to be exchanged into the transactions, but the data are encrypted using symmetric keys which are shared in advance between the sender and receiver through a secure channel.
The author suggests to use Multichain because it provides low overhead for the transactions handling.

ProvChain \cite{liang_provchain:_2017} is a blockchain based data provenance architecture for assuring data operation (i.e., data access and data changes) in the cloud storage application.
Different from the previous two studies, the major challenge is that the provenance data are also sensitive but still need to be validated by a third party. The authors proposed an additional layer as provenance auditor which interacts with a blockchain network by blockchain receipts which include provenance entry for future validation.


To the best of our knowledge, there is no study investigating the design of efficient logging and querying for a blockchain system at the application layer.  
The 2018 iDASH competition first track provides us with an opportunity to explore such possibilities.  
We attempt to design a blockchain-based log system that can serve as a light-weight and widely compatible component for the existing blockchain platforms.
Especially, our solution is optimized for genomic dataset access auditing under the requirements of the competition task.

\section*{Method}
We design a blockchain-based log system that is time/space efficient to store and retrieve genomic dataset access audit trail. Our method only leverages the Blockchain mechanism and is not limited to any specific Blockchain implementation, such as Bitcoin\cite{Bitcoin}, Ethereum\cite{Ethereum}. We introduce an on-chain indexing data structure which can be easily adapted to any blockchains that use a key-value database as their local storage. In our development, we use Multichain version 1.0.4 as an interface between Bitcoin Blockchain and our insertion and query method. 
Multichain is a Bitcoin Blockchain fork. It conveniently provides a feature, data stream, to allow us to use Bitcoin Blockchain as an append-only key-value database.


\subsection*{Overview}

In Figure \ref{fig:overview}, we illustrate the overview of the logging system, which is built on top of Multichain APIs.
The core task is to design space and time efficient methods for insertion and queries.
As described in Section ``Technical details of the task'', there are three types of primitive queries: point query, AND query, and range query. There are seven fields in the given genomic dataset: $\mathsf{Timestamp}$, $\mathsf{Node}$, $\mathsf{ID}$, $\mathsf{Ref-ID}$,  $\mathsf{User}$,  $\mathsf{Activity}$,  $\mathsf{Resource}$ as shown in Table \ref{smaple_logs}. 

\begin{table}[h!]
  \begin{tabular}{lllllll}
    \hline
    Timestamp & Node & ID & Ref-ID  &User& Activity  &Resource \\ \hline
    152202801 & 1& 1 &1& 1 &REQ\_RESOURCE& MOD\_FlyBase\\
    152208352 &1& 2 &1 &1& VIEW\_RESOURCE& MOD\_FlyBase\\
    152216966 &1 &3 &3 &6& FILE\_ACCESS&  GTEx\\
    152237149 &1&	9 &9	&10& REQ\_RESOURCE& MOD\_SGD \\ \hline
\end{tabular}
  \caption[Caption for LOF]{The Sample Logs.}
\label{smaple_logs}
\end{table}

For point query, the user can query on any field. 
For AND query, the user can query on any combination of fields. 
For range query, the user can query only on timestamp field with a start and end timestamp. See Table \ref{smaple_logs} \& \ref{insertion and queries examples} as a running example. 

\begin{table}[h!]
\begin{tabular}{l}
	\hline
	Insertion \\
         \tabitem \textit{Insert}(“152202801 1 1 1 1 REQ\_RESOURCE MOD\_FlyBase”)\\
    \hline
    Queries \\
        \tabitem \textit{Point\_Query}(Activity="VIEW\_RESOURCE") \\
        \tabitem \textit{AND\_Query}(ID="2", Node="1") \\
        \tabitem \textit{Range\_Query}(start=1522000000000, end=1522000100000)\\
        \hline
\end{tabular}
\caption{Insertion and Queries Examples.}
\label{insertion and queries examples}
\end{table}

\subsection*{Baseline method}

We first describe a naive method as a baseline. 
The baseline method leverages only three Multichain APIs as shown in Table \ref{t1}. 

\begin{table}[h!]
      \begin{tabular}{|p{5cm}|p{3cm}|}
        \hline
         Multichain APIs  & Description \\ \hline
        $\mathsf{create [stream\ name]}$ & reate a stream(table) in database  \\ \hline
        $\mathsf{publish [stream name] [key] [value]}$ & Insert key-value pair to specific stream(table)  \\ \hline
        $\mathsf{liststreamkeyitems [stream\ name] [key]}$ & Retrieve all items with the given key \\ \hline
      \end{tabular}
\caption{Multichain APIs Used in Our Methods.}
      \label{t1}
\end{table}

\noindent\textbf{Insertion:}
First, we create $K$ streams, where $K$ is the number of fields. \revA{Multichain builds $K$ tables in its back-end key-value database.} Second, we build $K$ key-value pairs, where the key is the attribute data and the value is entire record line. Finally, we convert those $K$ pairs into one Blockchain transaction and publish it to Blockchain. \revA{The following Figure \ref{Log Record to Transaction Conversion} is an example conversion from a log record to blockchain transaction. We will use this example log record in the remaining sections. After the transaction is confirmed by Blockchain, Multichain decodes the transaction and insert each key-value pairs to its corresponding table.}

\begin{figure}[h!]
  \centering
  \includegraphics[width=0.95\columnwidth]{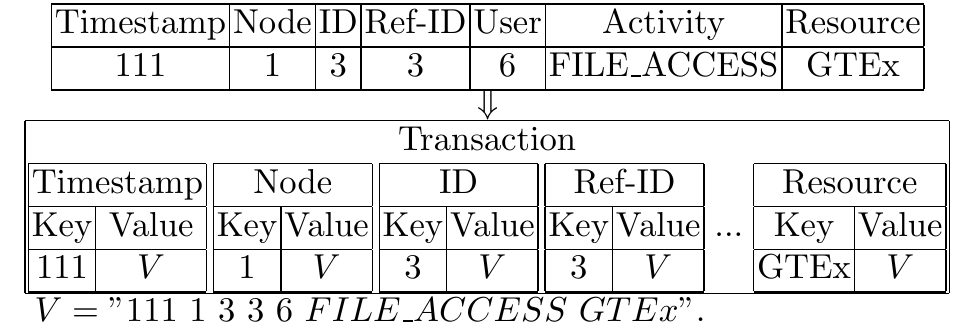}
  \caption{Log Record to Transaction Conversion.}
  \label{Log Record to Transaction Conversion}
\end{figure}

\noindent \textbf{Point Query:}
The implementation of a point query is straightforward which simply returns a list of records as shown in Algorithm \ref{algo:PQ}. In this literature, we assume the run time complexity of all Multichain API is $O(1)$. \revA{The run time complexity of Point Query is $O(1)$.}

\begin{algorithm}[h!]

		\SetKwRepeat{doWhile}{do}{while}
		\SetKwComment{tcc}{//}{}
		\DontPrintSemicolon 
		\KwIn{$A$, $K$ //attribute and key }
		\KwOut{$l_r$ //a list of record}

    	$l_r \gets $ \textit{liststreamkeyitems}($A$,$K$)
	    
		\Return{$l_r$}  
		\caption{Point Query}
		\label{algo:PQ}

\end{algorithm}

\noindent\textbf{AND Query:}
AND query enables a user to query with multiple keys.  We convert AND query to multiple point queries and intersect the result of all point queries. \revA{The run time complexity is
$O(K)$, where $K = number\ of\ keys$.}
\begin{algorithm}[h!]

		\SetKwRepeat{doWhile}{do}{while}
		\SetKwComment{tcc}{//}{}
		\DontPrintSemicolon 
		\KwIn{$l_{AK}$ // a list of attribute and key pairs}
		\KwOut{$l_r$ //a list of record}
		$l_r \gets \textit{point\_query}(l_{AK}[0]_A,l_{AK}[0]_K)$
		
        \ForEach{$(A,K) \in l_{AK}$}{
            $l_r \gets l_r \cap \  \textit{point\_query(A,K)}$
        }
		\Return{$l_r$}  
		\caption{AND Query}
		\label{algo:findSup}

\end{algorithm}

\noindent\textbf{Timestamp Range Query:}
Given a start timestamp and an end timestamp, Timestamp Range Query returns records whose timestamp is in this range.  We convert Timestamp Range Query into R point queries, where R is the range of timestamp. \revA{The run time complexity is $O(R)$, where $R = range\ of\ timestamp$.}
\begin{algorithm}[h]

		\SetKwRepeat{doWhile}{do}{while}
		\SetKwComment{tcc}{//}{}
		\DontPrintSemicolon 
		\KwIn{$t_s$, $t_e$ // start timestamp and end timestamp}
		\KwOut{$l_r$ //a list of record}
		
	    $l_r \gets \{\}$
		
		\For{$t=t_s\ \textbf{to}\ t_e$} {
	        $l_r \gets l_r\ \cup \ $\textit{point\_query}("Timestamp",t)
		}
		\Return{$l_r$}  
		\caption{Timestamp Range Query}
		\label{algo:findSup}

\end{algorithm}

\subsection*{Enhanced method}
After testing the baseline solution, which will be discussed in the result section, we found that the retrieve speed heavily depends on the number of API calls. Therefore, the fewer API calls we use, the faster retrieve speed we get. More specifically, we found three non-optimal issues:
\begin{itemize}
    \item The entire record is duplicated K times where K is the number of fields, which is insufficient in terms of storage overheads.
    \item Since we need to query all results and intersect them in local memory, AND query takes significant amount of memory when the number of AND operations increases. 
    \item  If the length of a given range query is $n$ (typically, $n$ is ranging from $10^6$ to $10^8$), the baseline method naively translate the range query into $n$ point queries and concatenate the results.
\end{itemize}
The blockchain-based auditing system is an append-only structure, so a data structure that keeps the minimum amount of information while maintaining the efficiency is essential. \revA{The percentage of read(query) operations in the real-world auditing system is low \cite{roselli1998characteristics}, therefore we trade retrieval speed for storage cost.} We redesign the key-value pairs in the blockchain transaction, modified the query algorithm accordingly and built a selectivity list based on data distribution. Most of all, we design a hierarchical timestamp structure which significantly reduces the number of queries(APIs) needed for the range query.\\
 
\begin{figure}[h!]
  \centering
  \includegraphics[width=0.95\columnwidth]{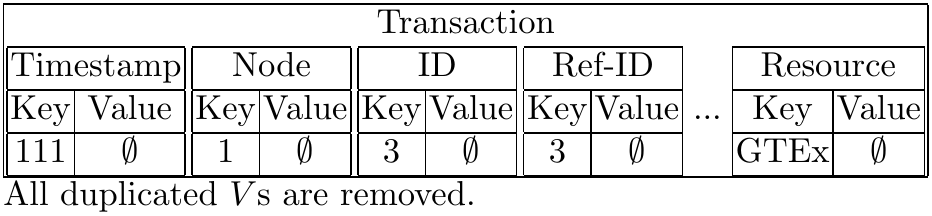}
  \caption{Transaction with empty values.}
  \label{Transaction with empty values}
\end{figure}

\noindent \textbf{Insertion:}
To address these problems, we redesign the key-value pairs. 
The key part remained the same (attribute data), but we removed the entire entry from the value part. As a result, we removed all duplicated values in the baseline method as shown in Figure \ref{Transaction with empty values}

\noindent \textbf{Point Query:}
Since we now have an empty value in the key-value database, we cannot use the key to get original record directly. We now take advantage of Blockchain transaction ID which is included in the returning JSON file of \textit{liststreamkeyitems} API. First, we get a list of \textit{TXID} (transaction ID) with the given key. Second, we use another Multichain API, \textit{getrawtransaction}, to get the matching transactions. Finally, we rebuild the original record from the transaction where all attribute data are included. It is worth mentioning that the point query now requires T + 1 times API calls to retrieve the records where T is the size of the \textit{TXID} list. Future study may able to combine those three steps into one if we modify Multichain version 1.0.4 source code, but it is not allowed in this competition. \revA{The run time complexity is $O(R)$, where $R = number\ of\ return\ record$.}

\begin{algorithm}[h]
		\SetKwRepeat{doWhile}{do}{while}
		\SetKwComment{tcc}{//}{}
		\DontPrintSemicolon 
		\KwIn{$A$, $K$ //attribute and key }
		\KwOut{$l_r$ //a list of record}

    	$[\textit{TXIDs}] \gets $ \textit{liststreamkeyitems}($A$,$K$)
    	
    	$l_r \gets []$
    	
    	\ForEach{$\textit{txid} \in [TXIDs]$}{
    	   $T \gets$ \textit{getrawtransaction} (\textit{txid})
    	   
    	   $R \gets \textit{rebuild(T)}$
    	   
    	   $\textit{append}(l_r, R)$
    	}
	    
		\Return{$l_r$}  
		\caption{Point Query with additional step}
		\label{algo1}
\end{algorithm}

\noindent \textbf{AND Query:} 
\revA{In order to reduce the retrieval cost,  we build a selectivity list for attributes based on the example test data which was given by the competition organizer. A selectivity list is based on the rank of result size of each attribute. The attribute which has the smallest query result size is the most selective. For the enhanced AND query, we call point query only one time for the most selective key then filter the result in the memory. Since we only query once from Blockchain, the total memory usage is bounded by the largest query result.} \revA{The run time complexity is $O(1)$}.

\begin{algorithm}[h]
		\SetKwRepeat{doWhile}{do}{while}
		\SetKwComment{tcc}{//}{}
		\DontPrintSemicolon 
		\KwIn{$l_{AK}, l_S$ // a list of attribute and key pairs and a selectivity list}
		\KwOut{$l_r$ //a list of record}
		$SK \gets \textit{findMostSelectiveKey}(l_{AK}, l_s)$
		
		$l_r \gets \textit{point\_query}(SK_A, SK_K)$

        \ForEach{$(A,K) \in l_{AK}$}{
            $l_r \gets \textit{filter}(l_r, A, K)$
        }
        
		\caption{AND Query with selectivity list}
		\label{algo:2}
\end{algorithm}

\noindent \textbf{Timestamp Range Query:}
Since Blockchain is an immutable structure, the common indexing techniques, \revA{such as B-tree and R-tree,} which require adjusting/balancing the entire data structure according to the data distribution, won’t work. We introduce a hierarchical timestamp structure, \revA{which is an incremental data structure and matches the append-only characteristics of the blockchain system.} Our design significantly reduces the number of queries(APIs) needed for a single range query.

\revA{The hierarchical timestamp structure consists of multiple levels. See Table \ref{Hierarchical Timestamp Structure} as an example. The range in the high level divides into multiple smaller range in the lower level. We denote each range part as LevelNumber:Starting Timestamp. A timestamp is recorded in the corresponding part at all levels. In our running example, a timestamp \textit{111} will be recorded in L0:100, L1:110, and L2:111 in Table \ref{Hierarchical Timestamp Structure}. }

\begin{table}[h!]
\begin{tabular}{c|c|c|c|c|c|c|c|c|c|}
L0 & \multicolumn{9}{c|}{{[}100,200)} \\ \cline{2-10} 
L1 & \multicolumn{3}{c|}{{[}100,110)} & \multicolumn{3}{c|}{{[}110,120)} & \multicolumn{3}{c|}{{[}120,..)} \\ \cline{2-10} 
L2 & 100 & ... & 109 & 110 & ... & 119 & 120 & ... & ... \\ \cline{2-10}
\end{tabular}
\caption{Simple Hierarchical Timestamp Structure }
\label{Hierarchical Timestamp Structure}
\end{table}

\begin{figure}[h!]
  \centering
  \includegraphics[width=0.95\columnwidth]{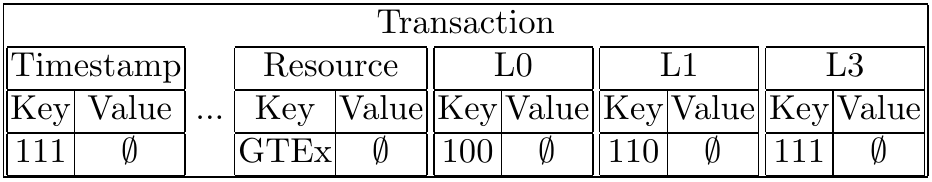}
  \caption{Transaction with Hierarchical Timestamp Structure.}
  \label{Transaction with Hierarchical Timestamp Structure}
\end{figure}

\revA{To build this structure, we need to slightly modify the insertion method by adding L streams where L is the number of levels, and we need to add L key-value pairs to Blockchain transaction as well. See Figure \ref{Transaction with Hierarchical Timestamp Structure} as an example.}

\revA{In our enhanced range query method, we recursively find the largest range in the hierarchical timestamp structure and use multiple point queries to retrieve the result.}

\begin{algorithm}[h!]

		\SetKwRepeat{doWhile}{do}{while}
		\SetKwComment{tcc}{//}{}
		\DontPrintSemicolon 
		\KwIn{$t_s$, $t_e$ // start timestamp and end timestamp}
		\KwOut{$l_r$ //a list of record}
		
	    $l_r \gets list$
	
		$l, r \gets \textit{findLargestRange}(t_s,t_e)$ 
		
		\While{$r \neq None$}{
		    $\textit{append}(l_r, \textit{point\_query}(l,r))$
		    
		    $l, r \gets \textit{findLargestRange}(t_s,t_e)$ 
		}  
		\Return{$l_r$}  
		\caption{Timestamp Range Query with hierarchical timestamp structure}
		\label{algo:findSup}
                
              \end{algorithm}

\revA{In the following example, we show the number of queries(APIs) needed for our baseline range query and enhanced range query.}

\begin{figure*}[h!]
    \centering
    \includegraphics[width=17cm]{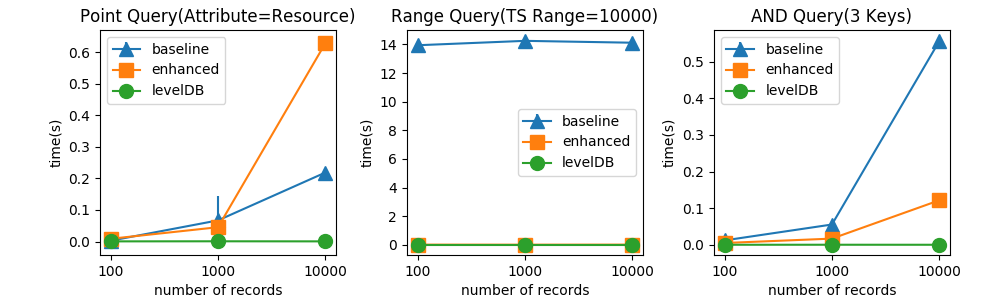}
    \caption{Scalability Test: Queries.}
    \label{Scalability Test: Queries}
\end{figure*}

\noindent \revA{Range query from timestamp $109$ to timestamp $120$.}

\noindent \revA{ Baseline Method:\\ $q('T',109) \cup q('T',110) \cup ... \cup q('T',120) \rightarrow$ 11 queries}

\noindent \revA{ Enhanced Method: \\ $q('L2',109) \cup q('L1',110) \cup q('L2',120)  \rightarrow$ 3 queries}

\revA{We reduce the number of queries needed for range query from $R_{T_e-T_s}$ to $\sum_{i=0}^L \frac{R_i}{r_{L_i}}$, where $R_{i+1} = R_i \mod r_{L_i},\ R_0 = R_{T_e-T_s}$ and $r_{L_i}$ is the elemental range at level $L_i$.}

\revA{The run time complexity of the enhanced range query is $O(\sum_{i=0}^L \frac{R_i}{r_{L_i}})$}.

\subsubsection*{Further optimizations}
The database normalization can be used for both baseline and enhanced solution. According to the given genomic datasets,
\textit{Ref-ID} refers back to the same original \textit{ID} which means those \textit{User} and \textit{Resource} are the same. For this reason, we can exclude \textit{User} and \textit{Resource} in Blockchain transaction.

\section*{Results}

\subsection*{Implementation environment}

We used Python3 as our main programming language to develop our solution, Savior \cite{Savoir} to interact with Multichain API and Docker \cite{Docker} to simulate 4 Blockchain nodes. Additionally, we created some bash scripts to automatically setup Blockchain nodes and Multichain environment. We also wrote a benchmark program to compare our baseline method and enhanced method. Our code is available online \cite{our-code}. The specifications of our testing machine are as follows: 6 cores CPU(i7 8700k), 32 GB of RAM and 6TB of HDD with Ubuntu 16.04 as the operating system. 

We used the sample testing data supplied by the competition organizer to benchmark our implementation. The sample testing data consists of 4 files, one per node. Each file has $10^5$ entries of log records which has 7 fields ($\mathsf{Timestamp}$, $\mathsf{Node}$, $\mathsf{ID}$, $\mathsf{Ref-ID}$,  $\mathsf{User}$,  $\mathsf{Activity}$,  $\mathsf{Resource}$). To illustrate, we provide a few sample data in Table \ref{smaple_logs}.

\revA{To find the optimal number of levels and the step multiplier of two adjacent levels for the hierarchical timestamp structure (Figure \ref{Hierarchical Timestamp Structure}), we test all reasonable parameter combinations by brute-force. For the given sample data, the optimal parameter for the number of levels is 3 and the step multiplier of two adjacent levels is 100. Future work may include finding the optimal number of levels in a more efficient way.}

\subsection*{Benchmark}
\revA{In our benchmark experiment, we show the scalability of our two methods alongside LevelDB\cite{LevelDB} as a reference, a popular key-value based database system which is used to store data in many blockchain systems\cite{androulaki_hyperledger_2018}\cite{Bitcoin}\cite{Ethereum}. Database system and Blockchain do not share the same design goal: the former is usually administered by a centralized entity, and the latter intents to work at a trustless environment. Nevertheless, this comparison offers useful insights of Blockchain based log system which trades speed for data integrity. We simulate the enhanced insertion, the enhanced point query, and the enhanced AND query behavior in LevelDB. For range query, we use LevelDB native method so we can properly examine our hierarchical timestamp structure. In all tests, we run 10 rounds for each methods with respect to varying the number of records. We calculate the average and the standard deviation from the results. We notice that the standard deviation is extremely small which shows the little trace in all figures expect Figure \ref{Scalability Test: Queries}(Point Query). This is due to the identical environment and the setup of our simulated blockchain nodes.}

\revA{\subsubsection*{Scalability Test: Queries}}
\revA{Figure \ref{Scalability Test: Queries} shows query time with respect to the varying number of records for point query, range query, AND query. For the point query test, the response time is determined by the result size. As the number of records increases, the result size increases and the response time increases. The response time of the enhanced method is worse than the baseline method because of the addition API calls which we introduced in the enhanced point query. For the range query test, the performance is constant since the result size of certain time range is constant. It is worth mentioning that our enhanced range query method have very close performance comparing to the native LevelDB range query method. For AND query, since it consists of point query, the response time increases with the increasing number of records. It is worth mentioning that the selectivity list design in our enhanced AND query method offsets the drawback of the enhanced point query method when the number of keys is larger than 2.}

\revA{\subsubsection*{Scalability Test: Insertion}}

\begin{figure}[h!]
  \centering
   \includegraphics[width=\columnwidth]{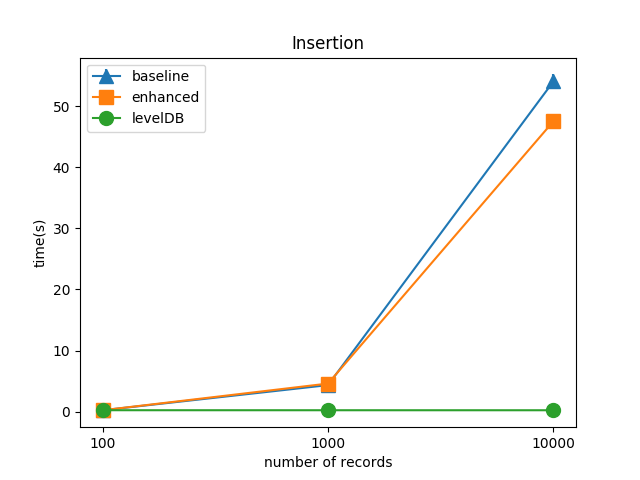}
  \caption{Scalability Test: Insertion.}
  \label{Scalability Test: Insertion}
\end{figure}

\revA{Figure \ref{Scalability Test: Insertion} shows the completion time of insertion methods with respect to varying the number of records. The insertion time is depended on the transaction size. The insertion times of the two methods are approximately the same. The enhanced method needs more key-value pairs to support hierarchical timestamp indexing structure. However, the empty values in key-value pairs offset this transaction size increment.}

\revA{\subsubsection*{Scalability Test: Storage}}

\begin{figure}[h!]
  \centering
  \includegraphics[width=\columnwidth]{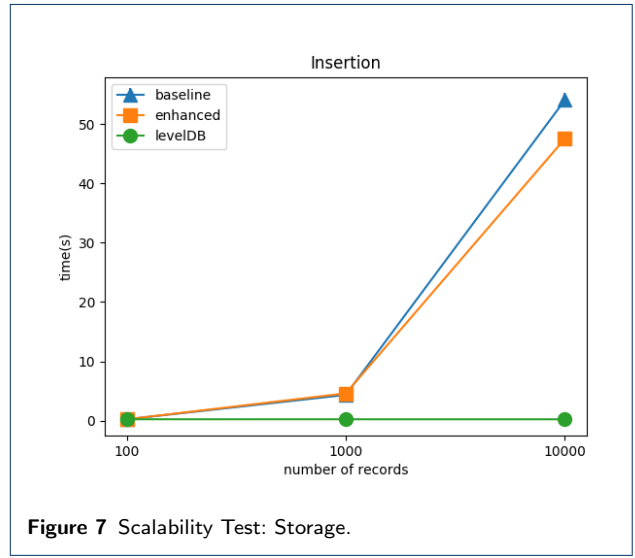}
  \caption{Scalability Test: Storage.}
  \label{Scalability Test: Storage}
\end{figure}

\revA{Figure \ref{Scalability Test: Storage} shows the total blockchain size in bytes with respect to varying the number of records. The blockchain size information is collected by calling Multichain API. Since Blockchain and LevelDB measure their size in different ways, we exclude LevelDB in this test. The figure suggests that the enhanced method uses less storage than the baseline method. The duplication removal from the blockchain transaction in the enhanced method works as designed. }

\revA{\subsubsection*{Detailed Comparison}}
\revA{In this section, we show a detailed performance difference of 3 query types in the baseline method, enhanced method, and LevelDB. We use the fixed 1000 records in the remaining tests. }

\begin{figure}[h!]
  \centering
   \includegraphics[width=\columnwidth]{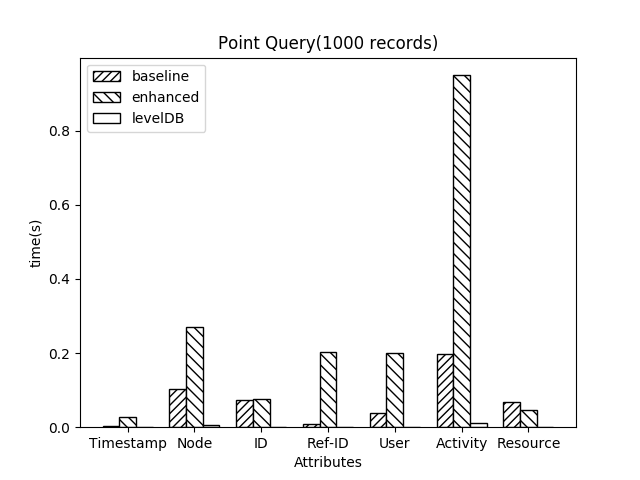}
  \caption{Point Query.}
  \label{Point Query}
\end{figure}

\revA{\noindent \textbf{Point Query:} Figure \ref{Point Query} shows the query response time for different attributes. The enhanced method performance is worse than the baseline method, because of the additional API calls in the enhanced method. The rank in the result also matches the rank in selectivity list which indicates the return record size. The return record size of \textit{Activity} is the largest among the attributes. In other words, \textit{Activity} has the lowest selective and need more API calls to get the result than other attributes, so it has the worst performance difference.}

\begin{figure}[h!]
  \centering
   \includegraphics[width=\columnwidth]{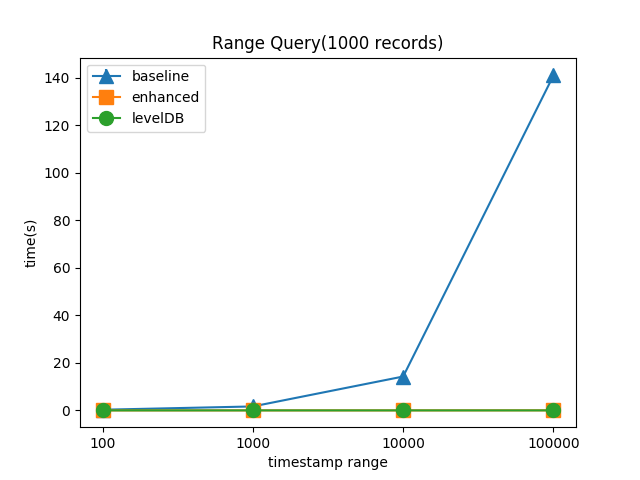}
  \caption{Range Query.}
  \label{Range Query}
\end{figure}

\revA{\noindent \textbf{Range Query:} Figure \ref{Range Query} shows the query response time with respect to varying the time range. The enhanced method is at least one order of magnitude better than the baseline method. Moreover, the Enhanced method has almost the same constant time performance as LevelDB native range query method. It proves that our hierarchical timestamp structure works as designed.}

\begin{figure}[h!]
  \centering
   \includegraphics[width=\columnwidth]{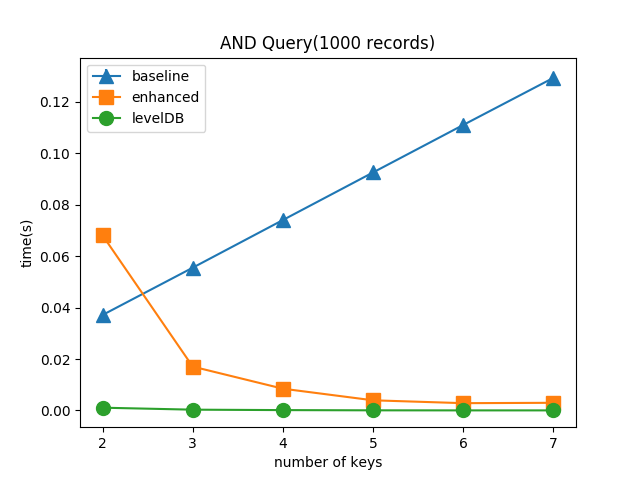}
  \caption{AND Query.}
  \label{AND Query}
\end{figure}

\revA{\noindent \textbf{AND Query:} Figure \ref{AND Query} shows the query time with respect to varying the number of keys. We test all combinations of keys. For example, for 2 keys test, we test all 21 combinations(7 choose 2) and average the result. It is much easier to find a more selective key when the number of keys is increasing. This is the reason why the enhanced method has a downward slope. When there are only 2 keys, the enhanced method has high possibility to find a low selective key. As a result, when AND query takes a low selective key, it requires a long response time.}

\section*{Discussion}
\revA{Our design is heavily governed by the competition requirements, evaluation criteria\cite{idash}, and Multichain 1.0.4 capability. In this paper, we intend to use Multichain as only an interface, so our design can be applied to any arbitrary blockchain system. Multichain 1.0.4 does not allow an item to have multiple keys and competition does not allow participators to modify Multichain, so we have to manually construct the blockchain transaction. There are two major developments for the future work: 1) A new interface which encodes/decodes the log entry to/from blockchain transaction more efficient. For example, in Bitcoin blockchain transaction script, it can write entire log entry only once in the blockchain transaction and let local interface client translate the script to the database. A specific Bitcoin interface for this log system can significantly reduce the transaction size. 2) A new Blockchain oriented database system, such as Forkbase\cite{wang_forkbase_2018}. It aims to design a new key-value architecture to reduce the development efforts of the above application and provide efficient analytical query performance. It is possible to replace the key-value engines in the existing blockchain platforms for better query performance.}

In this paper, we focus on designing efficient logging and querying schemes for immutable blockchain systems, and assume the blockchain network has been well-established under a specific \textit{consensus algorithm} and acceptable \textit{transaction throughput}.
In the following, we discuss how they may affect our solution.


Consensus algorithm may affect the performance of insertion functions because a newly generated access log (as a transaction) need to be accepted by all node in the network (achieving a consensus on the next block) in order to be stored in the ledger.
Consensus algorithm manifests the transaction throughput, which is majorly controlled by a predefined parameter in Multichain called \textsf{mining-diversity} (the default configuration is $0.3$).
If the transaction throughput is low, the insertion would be insufficient since it may be suspended until the previous batch of logs is finished.
The transaction throughput also affects the audit queries because the query is performed on the locally synchronized ledger.
Under low transaction throughput, a newly generated log may take a long time to be included in the ledger and synchronized to a node so that the query on a node may not be able to provide the accurate real-time answer.


Further, the access log could be private since it records all of the queries issued by a user.
This is a challenge for existing blockchain platforms since the ledger is public to every node in the network for increasing transparency and security.
A recent version of Hyperledger Fabric \cite{androulaki_hyperledger_2018} includes a new function for this problem.
The idea is dividing the ledger to different channels and selectively sharing a channel among a subset of users.
There are also other efforts for this problem by adopting secure multiparty computation \cite{zyskind_enigma:_2015},  zero-knowledge proof \cite{froelicher_unlynx:_2017} or trusted hardware \cite{hynes_demonstration_2018}.
Although this problem is beyond the scope of this competition, our solution could be extended using the above techniques.

\section*{Conclusions}
In this paper, we presented two solutions for blockchain-based logging and querying genomic dataset audit trail.  We built a baseline solution and then adjusted our implementation based on the evaluation criteria of the competition\cite{idash} and the general real-world characteristics of log systems\cite{Rosenblum:1992:DIL:146941.146943}. The blockchain-based log system is an append-only structure, so the storage increases monotonically. In the real world, the percentage of writing operation(insertion) is much higher than the portion of reading operation(query) in the workload  \cite{Rosenblum:1992:DIL:146941.146943}. Based on the above two reasons, we decided to prioritize the storage space over retrieval speed and insertion speed. We can reduce the storage cost by 25\% and increase the range query speed by at least one order of magnitude.  We claim that our hierarchical timestamp structure design is Blockchain implementation independent. It can be adapted to any Blockchain (e.g., Bitcoin, Ethereum, Hyperledger) with the help of an intermediary, such as Multichain.

\begin{backmatter}

\section*{Abbreviations}
GDPR: General data protection regulation; API: Application program interface; JSON: JavaScript object notation; TXID: Transaction identification; GTEx: Genotype-tissue expression

\section*{Ethics approval and consent to participate}
Not applicable.

\section*{Consent for publication}
Not applicable.

\section*{Availability of data and materials}
The data that support the findings of this study are available from the iDash workshop. The code is available at Github(https://github.com/mshuaic/Blockchain\_med/).

\section*{Competing interests}
The authors declare that they have no competing interests.
\section*{Author's contributions}
SM, YC and LX designed the solution and wrote the manuscript. 
SM implemented the code. All authors read and approved the final manuscript.



\section*{Acknowledgements}
We thank Dr. Tsung-Ting Kuo, the organizer of iDASH competition 2018 first track, for providing informative Q\&A and helpful advice!

\section*{Funding}
This work was supported by NIH R01GM118609, Georgia CTSA under grant NIH UL1TR002378, NSF under grant CNS-1618932, the AFOSR DDDAS program under grant FA9550-121-0240, the Japan Society for the Promotion of Science (JSPS) Grant-in-Aid for Scientific Research No. 17H06099, No. 18H04093 and No. 19K20269. The publication costs were funded by NIH R01GM118609, Georgia CTSA under grant NIH UL1TR002378, Japan Society for the Promotion of Science (JSPS) No. 17H06099, No. 18H04093, No. 19K20269.


\bibliographystyle{bmc-mathphys} 
\bibliography{idash18-citations}      

\end{backmatter}
\end{document}